# ASYMPTOTIC BOUNDS FOR SPHERICAL CODES

Yuri I. Manin[1], Matilde Marcolli[2]

[1] *Max–Planck–Institut für Mathematik, Bonn, Germany,*

[2] *California Institute of Technology, Pasadena, USA*

*ABSTRACT.* The set of all error–correcting codes $C$ over a fixed finite alphabet $\mathbf{F}$ of cardinality $q$ determines the set of code points in the unit square $[0,1]^2$ with coordinates $(R(C),\delta(C))$:= *(relative transmission rate, relative minimal distance).* The central problem of the theory of such codes consists in maximising simultaneously the transmission rate of the code and the relative minimum Hamming distance between two different code words. The classical approach to this problem explored in vast literature consists in the inventing explicit constructions of "good codes" and comparing new classes of codes with earlier ones.

Less classical approach studies the geometry of the whole set of code points $(R,\delta)$ (with $q$ fixed), at first *independently* of its computability properties, and only afterwords turning to the *problems of computability, analogies with statistical physics* etc.

The main purpose of this article consists in extending this latter strategy to domain of *spherical codes.*

## 1. Introduction: notation and summary

**1.1. Error–correcting discrete codes, their parameters and code points.** Consider a finite set, *alphabet* $\mathbf{F}$, of cardinality $q\geq 2$. An (unstructured) *code* $C$ is a non–empty subset $C\subset \mathbf{F}^n$ of words of length $n\geq 1$. Such $C$ determines its *code point* $P_C=(R(C),\delta(C))$ in the $(R,\delta)$–plane, defined by the formulas

$$\delta(C):=\frac{d(C)}{n(C)},\quad d(C):=\min\{d(a,b)\,|\,a,b\in C, a\neq b\},\quad n(C):=n,$$

$$R(C):=\frac{k(C)}{n(C)},\quad k(C):=\log_q \mathrm{card}(C), \tag{1.1}$$

where $d(a,b)$ is the Hamming distance

$$d((a_i),(b_i)):=\mathrm{card}\{i\in(1,\ldots,n)\,|\,a_i\neq b_i\}.$$

In the degenerate case card $C = 1$ we put $d(C) = 0$. We will call the numbers $k = k(C)$, $n = n(C)$, $d = d(C)$, *code parameters* and refer to $C$ as an $[n, k, d]_q$–code.

Among the simplest and most popular examples of *structured* codes are linear subspaces $C \subset \mathbf{F}_q^n$ where the alphabet $\mathbf{F}$ is now endowed with the structure of finite field. For much more details and viewpoints, see [TsfaVlaNo07], [ManMar11], [Man12].

In this paper, we will be mostly interested in the case $q = 2$ and unstructured codes.

**1.1.1. Asymptotic bounds for error–correcting codes.** *Fix* $q$ and consider the set of all points $P_C$ in the $(R, \delta)$–plane corresponding to $[n, k, d]_q$–codes. Denote by $U_q$ the closure of this set.

The basic theorem about its structure asserts the existence of a continuous function $\alpha_q$ of one variable such that $U_q$ is the union of its subset $R \leq \alpha_q(\delta)$ and a cloud of isolated code points lying in the region $R > \alpha_q(\delta)$. (Graph of the) function $\alpha_q$ is called *asymptotic bound.*

There is another characterisation of the asymptotic bound. Namely, slightly change the definition (1.1) of code points, replacing in it $k(C)$ by the integer part $[k(C)]$ so that the relative transmission rate $R(C)$ is replaced by a rational approximation to it $R_{rat}(C)$. Call a code point $(R_{rat}, \delta)$ a point of infinite (resp. finite) multiplicity, if the number of codes projecting to this point is infinite (resp. finite).

In [Man12], it was proved that the set of *all rational points* in $\mathbf{Q}^2 \cap [0, 1]^2$ lying below or on the asymptotic bound $R = \alpha_q(\delta)$ consists precisely of all code points *of infinite multiplicity.*

Similar results, with a priori different bounds, can be proved for certain structured codes, e. g. linear ones.

The proof of the existence of the asymptotic bound (see [Man12] and [ManMar11]) relies upon properties of *spoiling operations* on codes, which we review below.

**1.1.2. Spoiling operations for dicrete codes.** In this subsection, we will fix an $[n, k, d]_q$–code $C$ over alphabet $\mathbf{F}$ and introduce notation for codes that can be obtained from it by three classes of simple operations

*The first class of operations.* Consider a partial function $f : \mathbf{F}^n \to \mathbf{F}$ and $i \in \{1, \dots, n+1\}$. Let $C_1 := C *_i f \subset \mathbf{F}^{n+1}$ be given by

$$(a_1, \dots, a_{n+1}) \in C_1 \text{ iff } (a_1, \dots, a_{i-1}, a_{i+1}, \dots, a_{n+1}) \in C\,,$$

and $a_i = f(a_1, \dots, a_{i-1}, a_{i+1} \dots, a_n)$.

The code $C_1$ is an $[n+1, k, d]_q$ code when $f$ is a constant function.

*The second class of operations.* For the same $C$ and $i \in \{1, \dots, n\}$, define $C_2 := C*_i \subset \mathbf{F}^{n-1}$ by

$$(a_1, \dots, a_{n-1}) \in C_2 \text{ iff } \exists b \in \mathbf{F} \text{ such that } (a_1, \dots, a_{i-1}, b, a_i, \dots, a_{n-1}) \in C.$$

Then $C_2$ an $[n-1, k, d]_q$ code.

*The third class of operations.* Let $C_3 := C(a, i) \subset C \subset \mathbf{F}^n$ be given by

$$(a_1, \dots, a_n) \in C_3 \text{ iff } a_i = a.$$

Then $C_3$ a $[n, k', d']_q$ code where $k - 1 \le k' < k, d' \ge d$.

The way to use spoiling operations in order to derive properties of the closure of the set of the relevant code points starts with the remark that for growing $n$, new points obtained from an old one $c$ lie in the well controlled regions of the diminishing neighbourhoods of $c$. For more details, see sec. 2.7 below, [Man12] and [ManMar11].

**1.2. Spherical codes and their parameters.** We will now recall some basic facts about spherical codes and their relations with binary codes and with sphere packings, from [KaLe78] and [ConSlo99].

A *spherical code* consists of a finite set $X = \{x_1, \dots, x_k\}$ of points on the unit sphere in the Euclidean space $S^{n-1} \subset \mathbf{R}^n$. Writing each $x_i$ as a sequence of its coordinates in $\mathbf{R}^n$, we see that $n$ is similar to the block length of a discrete code, whereas the number $k$ is again the cardinality of code words/points.

The relevant *version of Hamming distance* between two code points $x, y$ is less obvious. We give here three numerical characteristics of essentially the same geometric notion, transition between which can be considered as “change of variable”.

(i) The angle $\varphi(x, y)$ between lines $(0, x)$ and $(0, y)$ in $\mathbf{R}^n$ normalised by $\varphi(x, y) \in [0, \pi]$.

(ii) The scalar product $(x, y)$.

(iii) The (unoriented) geodesic distance $\mathrm{dist}(x, y)$ between $x$ and $y$ in $S^{n-1}$.

The respective "change of variable" formulas are

$$\cos\varphi(x, y) = (x, y) = 1 - \frac{1}{2}\mathrm{dist}(x, y)^2. \tag{1.2}$$

Finally, although there is no obvious analog of $q$ for spherical codes, the following construction bridges binary discrete codes and spherical ones and suggests to accept for the latter the constant value $q = 2$.

Let $C$ be a binary $[n, k, d]_2$-code. By writing $\mathbf{F}$ as $\{\pm 1\}$ we can interpret its code words with a subset of vertices of $n$–dimensional cube centred at the origin of $\mathbf{R}^n$. By further normalising these vectors with the factor $n^{-1/2}$, we can identify the code words $c \in C$ with vertices of an $n$–cube centred at the origin in $\mathbf{R}^n$ and inscribed in the unit sphere $S^{n-1}$. Thus, a discrete binary code $C$ produces the spherical code $X_C$ of points on $S^{n-1}$, with the same parameters $n, k$.

The minimum Hamming distance $d$ of the binary code $C$ determines the minimal angle $\varphi$ between points of $X_C$ by

$$\cos\varphi = 1 - \frac{2d}{n} \tag{1.3}$$

Passing in the reverse direction, we have

$$\delta(C) = \frac{d}{n} = \sin^2(\varphi/2) = \frac{1 - \cos\varphi}{2}. \tag{1.4}$$

The transmission rate of the binary code is given by

$$R(C) = \frac{\log_2 \mathrm{card}\, X_C}{n} \tag{1.5}$$

Allowing in the latter formulas arbitrary spherical codes $X$ in place of $X_C$, we will from now on consider the following function $M(n, \varphi)$.

**1.2.1. Definition.** *$M(n, \varphi)$ is the maximal cardinality of $X \subset S^{n-1}$ satisfying any of the equivalent properties:*

*a) Spherical caps of angular radius $\varphi/2$ circumscribed around two different code points do not intersect.*

*b) The Euclidean distances between different code points are $\geq 2\sin(\varphi/2)$.*

*c) The scalar products between two different code points are $\leq \cos\varphi$.*

Starting from here, we will introduce in the next section spoiling operations and versions of asymptotic bounds for spherical codes.

**1.3. Plan of the paper.** The remaining main body of the paper consists of two sections.

In Section 2, we define a version of the set of spherical code points and various regions in their set related to the idea of Shannon optimality for information transmission via Gaussian channel with limited signal power. It leads to the introduction of the asymptotic boundary for spherical codes, and we prove several basic results about it.

In Section 3, we apply these results to sphere packings.

## 2. Code points and asymptotic bounds for spherical codes

**2.1. Code points and their domains.** For discrete (in particular, binary) codes, the domain accommodating all code points is the unit square $[0,1]^2$. of coordinates $(\delta, R)$ where the asymptotic bound $R = \alpha_q(\delta)$ lies.

For a spherical code $X$, as we argued in sec. 1.2, we can take as parameters the code rate $R = n^{-1}\log_2 \operatorname{card} X$ and the minimum angle $\varphi = \varphi_X$. Note that when $\varphi$ is sufficiently small, the maximal number of points $M(n, \varphi)$ on the sphere $S^{n-1}$ with minimal angle $\varphi$ correspondingly grows, hence the parameter $R$ is not a priori bound to be in the interval $[0,1]$ as in the case of binary codes.

Moreover, we will see that there are new phenomena that occur in the analysis of the asymptotic bound for spherical codes that do not happen in the case of binary and $q$–ary codes. These are due to the following basic fact. Imagine a code $X$ with very many code points and look at one point $x$ around which there are many other code points. "Generically", they will crowd in an $n-1$–dimensional subsphere of $S^{n-1}$ around $x$. But it might happen that neighbourhoods of smaller dimensions exist where most of these points lie.

To take this into account, we will have to introduce spoiling operations that depend on continuous parameters, and to give slightly different definitions of the regions we consider in the space of code parameters.

The space accomodating the set of code points $(R, \varphi)$ will be now $\mathbf{R}_+ \times [0, \pi]$. When convenient, we reparametrise the domain as $\mathbf{R}_+ \times [-1, 1]$ with coordinates $(R, \cos\varphi)$. As is customary in the codes literature, we plot $R$ along the vertical dimension and $\varphi$ along the horizontal dimension, even though we write the coordinates as $(R, \varphi)$.

**2.1.1. Definition.** *In the space $\mathbf{R}_+ \times [0, \pi]$ we define the following subsets.*

*a) The set of points that are code parameters of some spherical code $X$,*

$$\mathcal{P} = \{P = (R, \varphi) \,|\, \exists X \subset S^{n-1} \,:\, (R, \varphi) = (R(X) := \frac{1}{n} \log_2 \operatorname{card} X, \varphi_X)\}. \quad (2.1)$$

*b) The set of points that are accumulation points of code parameters*

$$\mathcal{A} = \{P = (R, \varphi) \,|\, \exists (R_i, \varphi_i) \in \mathcal{P} \,:\, (R, \varphi) = \lim_i (R_i, \varphi_i),\, (R_i, \varphi_i) \neq (R, \varphi)\}. \quad (2.2)$$

*c) The set of points surrounded by a ball densely filled by code parameters*

$$\mathcal{U} = \{P = (R, \varphi) \,|\, \exists \varepsilon > 0 \,:\, B(P, \varepsilon) \subset \mathcal{A}\}, \quad (2.3)$$

*where $B(P, \varepsilon)$ is the Euclidean ball of radius $\varepsilon$ around $P$ in $\mathbf{R}_+ \times [0, \pi]$.*

*d) The asymptotic bound $\Gamma$ is the set*

$$\Gamma = \{(R = \alpha(\varphi), \varphi) \,|\, \alpha(\varphi) = \sup\{R \in \mathbf{R}_+ \,:\, (R, \varphi) \in \mathcal{U}\}\}, \quad (2.4)$$

*where $\alpha(\varphi) = 0$ if $\{R \in \mathbf{R}_+ \,|\, (R, \varphi) \in \mathcal{U}\} = \emptyset$.*

One of the new features of the case of spherical codes, that does not occur in the case of discrete codes, is the fact that the two regions $\mathcal{A}$ and $\mathcal{U}$ do not coincide. The asymptotic bound we consider in this setting is the boundary of the region $\mathcal{U}$. As we discuss below, the part of the region $\mathcal{A}$ that is not in $\mathcal{U}$ consists of sequences of horizontal segments that are not contained in the set $\mathcal{U} \cup \Gamma$.

**2.1.2. The large angle range.** There are two separate regions with very different behavior in the analysis of spherical codes: the "small angle range", consisting of the set of spherical codes with minimum angle $0 \leq \varphi \leq \pi/2$, and the "large angle range" with $\pi/2 < \varphi \leq \pi$.

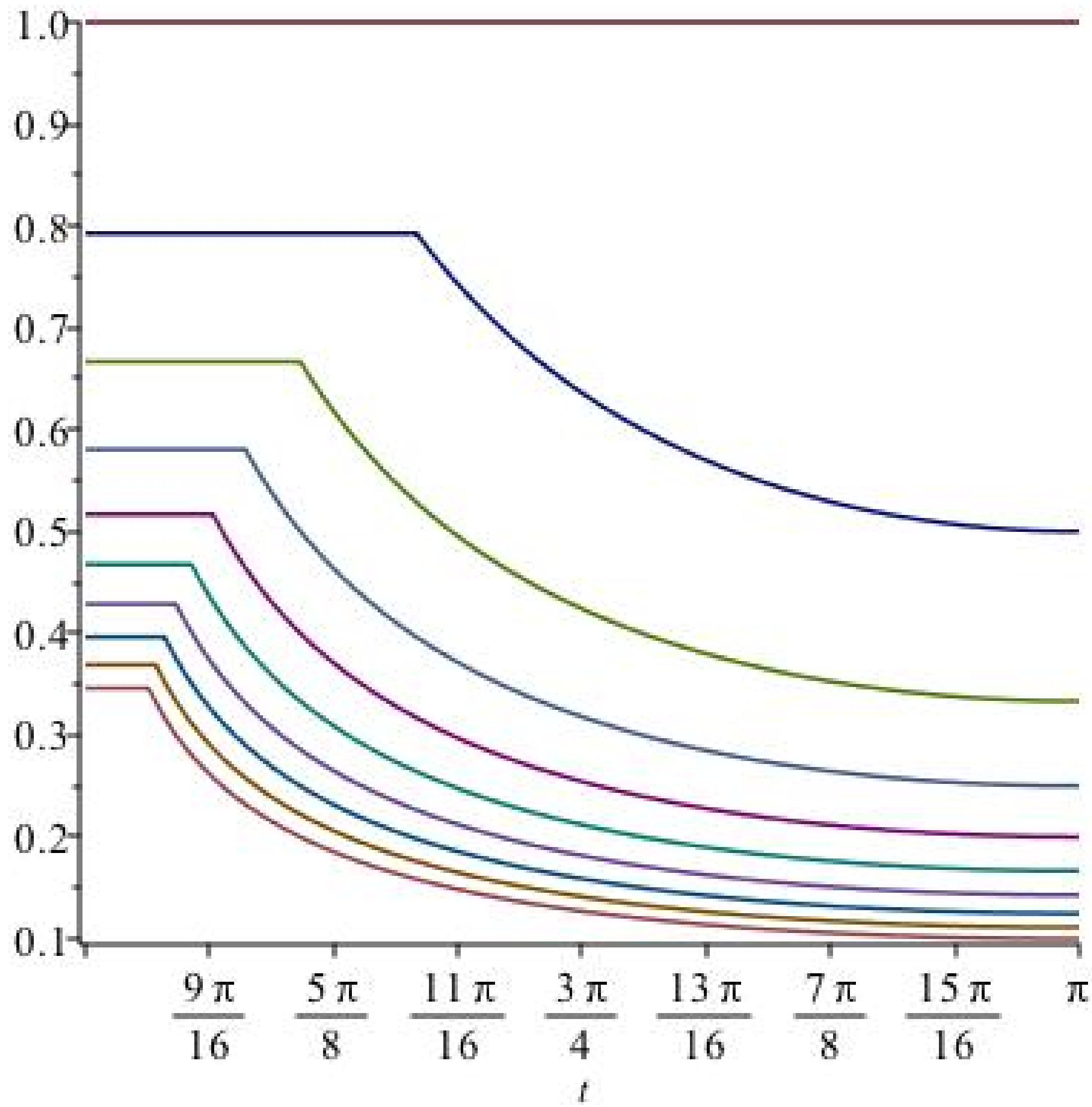
1.0
0.9
0.8
0.7
0.6
0.5
0.4
0.3
0.2
0.1
9π/16
5π/8
11π/16
3π/4
13π/16
7π/8
15π/16
π
t

**FIG. 1**

The results of [Ra55] (see also sec. 6 of [KaLe78]) show that for large angles $\pi/2 < \varphi \leq \pi$, the maximal number of points $M(n,\varphi)$ on the sphere $S^{n-1}$ with minimal angle $\varphi$ is bounded above by $(\cos\varphi - 1)/\cos\varphi$. The bound is realized for $-1 \leq \cos\varphi \leq -1/n$, while for $-1/n \leq \cos\varphi < 0$ one has $A(n,\varphi) = n+1$. Thus, in the large angle region the code points of spherical codes $X \subset S^{n-1}$ lie below the curve

$$R = \frac{1}{n}\log_2(\min\{n+1, \frac{\cos\varphi - 1}{\cos\varphi}\}).$$

These lines for varying $n = 1, \dots, 10$ are plotted in the Figure 1.

This implies that the large $n$ behaviour in this region gives

$$R = \frac{\log_2 \operatorname{card} X}{n} \leq \frac{\log_2 M(n,\varphi)}{n} \to 0, \qquad \pi/2 \leq \varphi \leq \pi$$

for $n \to \infty$, hence there is no interesting asymptotic bound in the large angle region.

However, as we discuss below, this large angle region still contributes code points in $\mathcal{A} \smallsetminus \mathcal{U}$ and in $\mathcal{P} \smallsetminus \mathcal{A}$.

**2.2. Code parameters and bounds in small angle range.** We now consider spherical codes that have minimum angle $0 \leq \varphi \leq \pi/2$. As we recalled above, binary codes $C$ determine associated spherical codes $X_C$, with parameters $k = \log_2 \operatorname{card} X_C$, $R = n^{-1}\log_2 \operatorname{card} X_C$ and $\delta = d/n = \sin^2(\varphi/2)$, which belong to this small angle region.

In particular, this implies that any upper bound for code parameters of spherical codes in the small angle range implies an upper bound on binary codes (but not vice versa, as not all spherical codes can be realised by binary codes).

In the small angle region, there is a linear programming upper bound for $M(n,\varphi)$, obtained in [KaLe78]. It gives the Kabatiansky–Levenshtein bound on code parameters $R$ for spherical codes, for $n \to \infty$ given by

$$R \leq \frac{\log_2 M(n,\varphi)}{n} \leq \frac{1+\sin\varphi}{2\sin\varphi}\log_2\left(\frac{1+\sin\varphi}{2\sin\varphi}\right) - \frac{1-\sin\varphi}{2\sin\varphi}\log_2\left(\frac{1-\sin\varphi}{2\sin\varphi}\right) \tag{2.5}$$

for minimum angle $0 \leq \varphi \leq \pi/2$. Thus, the space of code parameters of spherical codes, for large $n \to \infty$ and for small minimum angle $0 \leq \varphi \leq \pi/2$, is given by the undergraph

$$\mathcal{S} := \{(R,\varphi) \in \mathbf{R}_+ \times [0,\pi] \,:\, R \leq H(\varphi)\}, \tag{2.6}$$

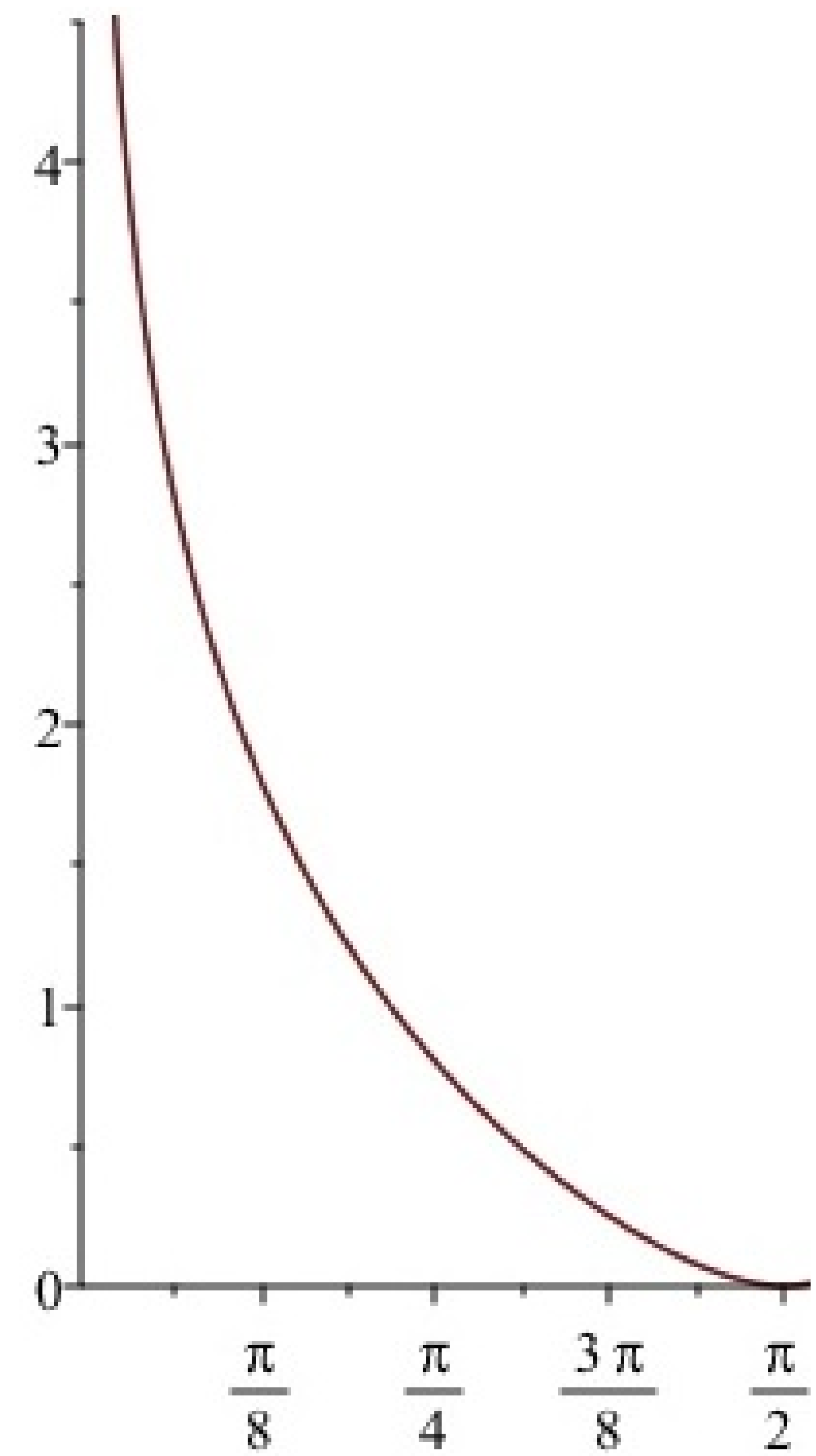

**FIG. 2**

$$H(\varphi) = \frac{1+\sin\varphi}{2\sin\varphi}\log_2\left(\frac{1+\sin\varphi}{2\sin\varphi}\right) - \frac{1-\sin\varphi}{2\sin\varphi}\log_2\left(\frac{1-\sin\varphi}{2\sin\varphi}\right). \tag{2.7}$$

In particular, the function $H(\varphi)$ diverges for $\varphi \to 0$ and does not provide any upper bound on the parameter $R$ of spherical codes, cf. Figure 2. This indeed corresponds to the fact that when the minimum angle $\varphi \to 0$ the number of points $\operatorname{card} X$ of the code can grow arbitrarily large, for any fixed $n$, resulting in an unbounded code parameter $R$. To avoid this problem, we will be considering a cutoff in this region.

One possibility is to consider an a priori cutoff on the minimum angle, by only considering spherical codes with $\varphi \geq \varphi_0$ for a chosen $\varphi_0 > 0$. This is a natural choice, for example, when focusing on spherical codes that are generated by sphere packings, for which there is a lower bound on the minimal angle $\varphi \geq \pi/3$.

Another possibility, which appears more natural with respect to the spoiling operations we discuss below, consists of introducing an a priori cut off on the parameter $R$ by considering only codes with bounded

$$R = \frac{1}{n}\log_2 \operatorname{card} X \leq T$$

for some fixed $T > 0$, that is, codes $X \subset S^{n-1}$ with number of points bounded by $\operatorname{card} X \leq 2^{nT}$.

We discuss in the Lemma 2.9.1 below how the asymptotic bound for spherical codes is related to the Kabatiansky–Levenshtein bound (2.5).

**2.3. Spoiling operations for spherical codes.** We consider the effects of the spoiling operations for binary codes on the parameters of the associated spherical codes, and we generalise these operations to a family of spoiling operations (which depends on continuous parameters) on the set of all spherical codes, not just those that come from binary codes.

For binary codes, the minimum angle satisfies $\cos(\varphi) = 1 - 2d/n$ (cf. (1.3)), hence the small angle range $0 \leq \varphi \leq \pi/2$ corresponds to the code parameter $\delta = d/n < 1/2$.

**2.4. The first class of operations for spherical codes.** When we associate to a binary code $C$ over the alphabet $\mathbf{F} = \{0,1\}$ the corresponding spherical code $X_C$ (cf. sec. 1.2), we can reinterpret the first spoiling operation $C_1 = C \star_i a$, which associates to a word $c = (a_1, \dots, a_n)$ of $C$ the word $c \star_i a = (a_1, \dots, a_{i-1}, a, a_i, \dots, a_n)$ of $C_1$, as the operation that takes the code $X_C \subset S^{n-1}$ and inserts this $S^{n-1}$ as a hyperplane section of the unit sphere $S^n$ of $\mathbf{R}^{n+1}$, where the hyperplane is given by

$$x_i = 1/\sqrt{n+1} \quad \text{if} \quad a = 0,$$

$$x_i = -1/\sqrt{n+1} \quad \text{if} \quad a = 1.$$

The resulting embedding of $X_C$ in $S^n$ gives a spherical code of dimension $n+1$, which is the spherical code $X_{C_1}$ associated to the spoiled code $C_1$.

The radius $\rho$ of the sphere $S^{n-1}_\rho$ cut out as the section of the unit sphere $S^n$ by the hyperplane $x_i = \pm 1/\sqrt{n+1}$ is given by $\rho^2 = 1 - \frac{1}{n+1}$. If $v_\ell \neq v_r$ are the vectors on the unit sphere $S^{n-1}$ corresponding to two points of the code $X_C$ with angle $\langle v_\ell, v_r \rangle = \cos\theta$, then the corresponding vectors in the spherical code $X_{C_1}$ are given by $\tilde{v}_\ell$ and $\tilde{v}_\ell$ where $\tilde{v}_\ell = \rho v_\ell^{(i)} + w$, with $\rho$ the scaled radius as above, $w$ the vector with coordinates $x_i = \pm 1/\sqrt{n+1}$ and $x_j = 0$ for $j \neq i$, and $v_\ell^{(i)} = v_\ell \star_i 0$ with the notation used above. Since the vector $w$ is orthogonal to the vectors $v_\ell^{(i)}$, the respective angle is given by

$$\langle \tilde{v}_\ell, \tilde{v}_r \rangle = \cos\tilde{\theta} = \rho^2 \langle v_\ell^{(i)}, v_r^{(i)} \rangle + \langle w, w \rangle = \frac{n}{n+1}\cos\theta + \frac{1}{n+1}.$$

The minimum angle for $X_{C_1}$ therefore satisfies

$$\cos\tilde{\varphi} = \frac{n}{n+1}\cos\varphi + \frac{1}{n+1} = \frac{n}{n+1}\left(1 - \frac{2d}{n}\right) + \frac{1}{n+1} = 1 - \frac{2d}{n+1},$$

since for the minimum angle for $X_C$ we have $\cos\varphi = 1 - \frac{2d}{n}$. This shows that the minimum distance is unchanged by this spoiling operation: $d = d(C) = d(C_1)$.

We will now extend this type of spoiling operation to more general spherical codes that do not necessarily arise from binary codes. In this general setting, however, the spoiling operation will depend on continuous parameters, unlike the case of binary codes, where it depends on the finite choice of $i \in \{1, \dots, n\}$ and $a \in \{0, 1\}$.

Let $H$ be an arbitrary hyperplane in $\mathbf{R}^{n+1}$ that intersects the unit sphere $S^n$ in more than one point, that is, in a sphere $S^{n-1}_\rho$. If we want the resulting sphere $S^{n-1}_\rho = H \cap S^n$ to have radius $\rho$ strictly less than 1, we further require that the hyperplane $H$ does not contain the origin.

**2.4.1. Definition.** *Given a spherical code $X \subset S^{n-1}$ and a hyperplane $H$ as above, the spoiling operation $X_1 := X \star H$ is obtained by scaling the sphere $S^{n-1}$ and identifying it with the section $H \cap S^n$. This gives an embedding of the set of points $X$ in the unit sphere $S^n$. The resulting set of points in $S^n$ is the spherical code $X_1$.*

As in the previous discussion we can see the effect of this spoiling operation on the code parameters of the spherical codes.

**2.4.2. Lemma.** *Let $X_1 = X \star H$ be the spoiled spherical code. Then $k(X_1) = k(X)$, $n(X_1) = n(X) + 1$, and the minimal angle $\varphi_{X_1}$ satisfies*

$$\cos \varphi_{X_1} = \rho_H^2 \cos \varphi_X + (1 - \rho_H^2), \tag{2.8}$$

*where $\rho_H$ is the radius of the sphere $S^{n-1}_\rho = H \cap S^n$.*

**Proof.** The parameter $k = \log_2 \operatorname{card} X = \log_2 \operatorname{card} X_1$ is unchanged, while $n \mapsto n + 1$ and the minimal angles are related by the same computation shown above. Namely, let $w$ be the vector in $\mathbf{R}^{n+1}$ orthogonal to the hyperplane $H$ (with length the distance of $H$ from the origin). Let $\rho = \rho_H$ be the radius of the sphere $S^{n-1}_\rho = H \cap S^n$. Given vectors $v_\ell, v_r \in X \subset S^{n-1}$ consider the corresponding vectors $v_\ell^{(H)}$ and $v_r^{(H)}$ in $\mathbf{R}^{n+1}$, which in a system of coordinates where $w$ is one of the axes, have zero coordinate along $w$ and the same coordinates as $v_\ell$ and $v_r$ along the other coordinate axes. The angles are related by

$$\cos \tilde{\theta} = \langle \rho_H \, v_\ell^{(H)} + w, \rho_H \, v_r^{(H)} + w \rangle = \rho_H^2 \, \cos \theta + (1 - \rho_H^2),$$

hence under the spoiling operation $C_1 = C \star H$ the minimal angle satisfies

$$\cos \varphi \mapsto \rho_H^2 \cos \varphi + (1 - \rho_H^2).$$

If $\rho_H$ is close to 1 (that is, the hyperplane section is close to the origin) then the minimal angle of $X_1$ is close to the minimal angle of the unspoiled spherical code $X$, while if $\rho_H$ is very close to zero (the hyperplane is close to being tangent to the sphere), then the cosine of the minimal angle of $X_1$ is very close to 1, hence the minimal angle of the spoiled spherical code $X_1$ becomes very close to zero.

**2.5. The second class of spoiling operations for spherical codes.** When the second spoiling operation is applied to a binary code $C$, it produces the code $C_2 = C\star_i$, which is the projection of the code $C$ in the $i$–th direction. Geometrically, this means the projection of the $n$–cube onto an $(n-1)$–cube in the coordinate hyperplane $x_i = 0$.

The sphere $S^{n-1}$ circumscribed around the $n$–cube is then projected onto the unit ball $B^{n-1}$ in this hyperplane. Since all the code points in the spherical code $X_C$ lie at vertices of the $n$-cube, none of them is mapped to the origin under this projection. Normalising the resulting projected vectors in $\mathbf{R}^{n-1}$ by he factor $\sqrt{n}/\sqrt{n-1}$, we get a new set of vectors in $\mathbf{R}^{n-1}$ corresponding to points on the sphere $S^{n-2} = \partial B^{n-1}$. The spherical code $X_{C_2} \subset S^{n-2}$ is obtained as the image of the points in the spherical code $X_C$ under this projection and rescaling.

If $C$ is a $[n,k,d]_2$–code, then $C_2 = C\star_i$ has code parameters $[n-1,k,d-1]$, provided that the projection (the letter place $i$) is chosen so that there are two words realizing the minimum distance $d$ that differ at the $i$–th letter. Otherwise $d$ will remain unchanged. Note that, if we associated a spherical code $X_C$ to $C$ as above, the change $n \mapsto n-1$, $k \mapsto k$, $d \mapsto d-1$ corresponds to changing $R \mapsto \frac{n}{n-1}R$ and $\delta \mapsto \frac{n}{n-1}\delta - \frac{1}{n-1}$, which in turn implies

$$\cos\varphi' = 1 - 2\delta' = 1 - 2\left(\frac{n}{n-1}\delta - \frac{1}{n-1}\right) = 1 - 2\frac{n}{n-1}\frac{1-\cos\varphi}{2} + \frac{2}{n-1}$$

$$= 1 + \frac{n}{n-1}\cos\varphi - \frac{n}{n-1} + \frac{2}{n-1} = \frac{n}{n-1}\cos\varphi + \frac{1}{n-1}.$$

Thus, applying the spoiling operation to the spherical code $X_C$ with code parameters $[n,k,\varphi]$ we obtain a spherical code $X_{C\star_i}$ with code parameters $[n-1,k,\varphi']$ where $\cos\varphi' \geq \cos\varphi$ is given by

$$\cos\varphi' = \frac{n}{n-1}\cos\varphi + \frac{1}{n-1}.$$

Furthermore, let $\cos\theta = \langle v_k, v_r\rangle$ be the angle between two points in the spherical code $X_C$, and $v_k^{\perp_i}$, $v_r^{\perp_i}$ denote their orthogonal projections along the $x_i$ axis, so that $\langle v_k, v_r\rangle = \langle v_k^{\perp_i}, v_r^{\perp_i}\rangle + \langle v_{k,i}, v_{r,i}\rangle$, with $v_{k,i}$ and $v_{r,i}$ the $i$–th component of the vectors.

The condition that the code words differ at the $i$–th letter (which is needed to lower the parameter $d$) corresponds to $v_{k,i}$ and $v_{r,i}$ having opposite signs, so the angle between the respective points in $X_{C_2}$ is given by

$$\cos\tilde{\theta} = \frac{n}{n-1}\langle v_k^{\perp_i}, v_r^{\perp_i}\rangle = \frac{n}{n-1}(\cos\theta - \langle v_{k,i}, v_{r,i}\rangle).$$

Since all the components of the vectors $v_k, v_r$ are equal to $\pm 1/\sqrt{n}$, but we are looking at two vectors for which they have different signs, we obtain $\langle v_{k,i}, v_{r,i}\rangle = -1/n$. Thus, we obtain finally

$$\cos\tilde{\theta} = \frac{n}{n-1}\cos\theta + \frac{1}{n-1}.$$

The minimum angle in $X_{C_2}$ then satisfies

$$\cos\varphi' = \frac{n}{n-1}(\cos\varphi + \frac{1}{n}) \geq \cos\varphi.$$

Motivated by these results, we will now define an analog of the second spoiling operation for general spherical codes $X \subset S^{n-1}$ .

**2.5.1. Definition.** *Let $L$ be an arbitrary hyperplane passing through the origin in $\mathbf{R}^n$ such that the line $\ell$ through the origin orthogonal to $L$ does not contain any point of $X$. Consider the orthogonal projection $P_L : \mathbf{R}^n \to L \simeq \mathbf{R}^{n-1}$ and the image $P_L(X) \subset B^{n-1} \smallsetminus \{0\}$. The subset $X_2 \subset S^{n-2}$ obtained by normalizing the vectors in $P_L(X)$ is the spherical code $X_2 = X\star_L$ determined by the spoiling operation.*

The effect of the second spoiling operation of spherical codes on the code parameters is as follows.

**2.5.2. Lemma.** *Let $X_2 = X\star_L$ be the spoiled spherical code. Then we have*

*a) $k(X_2) = k(X)$ and $n(X_2) = n(X) - 1$.*

*b) If the hyperplane $L$ is chosen so that there is a pair of vectors in $X$ realising the minimum angle $\varphi_X$ and the minimum distance of $X$ to $\ell$, with projections onto $\ell$ of opposite signs, then the minimal angle $\varphi_{X_2}$ satisfies*

$$\cos\varphi_{X_2} = (1+u)\cos\varphi_X + u, \qquad (2.9)$$

*where $u \geq 0$ is $u = (1 - \xi^2_{X,L})/\xi^2_{X,L}$, and $\xi_{X,\ell} := \mathrm{dist}(X, \ell)$ is the distance of $X$ from the line $\ell$.*

*Moreover, if the line $\ell$ bisects the minimum angle, then the minimal angle $\phi_{X_2}$ satisfies*

$$\cos \varphi_{X_2} = (1 + u) \cos \varphi_X - u, \tag{2.10}$$

*with $u$ as above.*

**Proof.** For a general $L$, the number of points $\mathrm{card}\, X = \mathrm{card}\, X_2$. This means that the transmission rate of the code $X_2$ is

$$R(X\star_L) = \frac{1}{n-1} \log_2 \mathrm{card} X\star_L = \frac{n}{n-1} R(X).$$

To compute the change in the minimum angle, we have as before

$$\cos \tilde{\theta} = \frac{\langle v_k^{\perp_L}, v_r^{\perp_L} \rangle}{\| v_k^{\perp_L} \| \cdot \| v_r^{\perp_L} \|} = \frac{1}{\| v_k^{\perp_L} \| \cdot \| v_r^{\perp_L} \|} (\cos \theta - \langle v_{k,\ell}, v_{r,\ell} \rangle),$$

where $v_{k,\ell}$ is the component along the line $\ell$ and $v_k^{\perp_L}$ is the orthogonal projection onto $L$, for $v_k, v_r \in X$ with $\langle v_k, v_r \rangle = \cos \theta$. The component $v_{k,\ell}$ and the vector $v_k^{\perp_L}$ satisfy the relation

$$\| v_k^{\perp_L} \|^2 + v_{k,\ell}^2 = 1.$$

Setting $x = \| v_k^{\perp_L} \|$ and $y = \| v_r^{\perp_L} \|$, we write the above as

$$\cos \tilde{\theta} = \frac{\cos \theta \mp \sqrt{1 - x^2} \cdot \sqrt{1 - y^2}}{x \cdot y},$$

where the sign depends on whether the two vectors lie in the same or in opposite hemispheres with respect to $L$.

The range of variability of $x, y$ can be visualized as follows. Its lower bound corresponds the to minimum value $\xi = \xi_{X,L} = \mathrm{dist}(X, \ell)$ given by the minimum of the distances of code points in $X$ to the line $\ell$, and a maximum possible value equals to one. The new minimum angle $\cos \varphi_{X_2} \geq \cos \varphi_X$ satisfies also the inequality

$$\cos \varphi_{X_2} \geq \frac{1}{\xi^2_{X,L}} (\cos \varphi_X + (1 - \xi^2_{X,L})).$$

This estimate can be achieved, for instance, if $L$ is taken so that there is a pair of vectors in $X$ realizing the minimum angle $\varphi_X$ that also have the minimal distance from $\ell$, and their projections along $\ell$ have opposite signs.

We write the above equivalently as $\cos\varphi_{X_2} = (1+u)\cos\varphi_X + u$ with $u = (1-\xi^2_{X,L})/\xi^2_{X,L}$. If $\ell$ is chosen to be the line that bisects the minimum angle in the plane containing two vectors of $X$ with the minimum angle, then the norm of the projections to $L$ of the vectors is $\xi_{X,L} = \sin(\varphi_X/2)$. In this case the $\ell$ projections of these two vectors have the same sign, so the resulting angle is

$$\cos\varphi_{X_2} = (1+u)\cos\varphi_X - u,$$

with $u = (1-\xi^2_{X,L})/\xi^2_{X,L}$ as above. In this case, $\cos\varphi_{X_2} \le \cos\varphi_X$.

**2.6. The third class of spoiling operations for spherical codes.** When we perform the third spoiling operation $C_3 = C(a,i)$ on a binary code $C$ with alphabet $\mathbf{F}_2$, we pass from $C$ to the subset of its words with letter $a$ as the $i$–th digit. For each $i$, it is always possible to find an $a \in \{0,1\}$ so that $2^{k-1} \le \operatorname{card} C(a,i) < 2^k$.

This corresponds, for the associated spherical code $X_C$, to taking the subset $X_{C_3}$ of points of $X_C$ that lie on the intersection of $S^{n-1}$ and the hyperplane $x_i = 1/\sqrt{n}$ for $a = 0$, and $x_i = -1/\sqrt{n}$ for $a = 1$. It also corresponds to taking the subset of points of $X_C$ that lie in the hemisphere with $x_i > 0$, or respectively $x_i < 0$.

For a general spherical code $X \subset S^{n-1}$, we generalise the third spoiling operation in the following way, which will provide the analog of the property $2^{k-1} \le \operatorname{card} C(a,i) < 2^k$ for binary codes.

Given our oriented line $\ell$ through the origin of $\mathbf{R}^n$ and the hyperplane $L$ through the origin orthogonal to $\ell$, let $S^{n-1}_{\ell,\pm}$, denote the two hemispheres corresponding to the non-negative and the negative half-spaces $\mathbf{R}^n_\pm$, with respect to the positive and negative parts of the coordinate line $\ell$.

**2.6.1. Definition.** *Let $\ell$ and $L$ be a line and the orthogonal hyperplane through the origin as above. The spoiled spherical codes $X_3 := X^\pm_\ell$ are given by the intersection of the code $X$ with one of the two hemispheres $X^\pm_\ell = X \cap S^{n-1}_{\ell,\pm}$.*

The effect of the third spoiling operations on the parameters of the spherical code is described as follows.

**2.6.2. Lemma.** *For any spherical code $X \subset S^{n-1}$, there is a choice of $\ell$ such that the spoiled code $X_3 = X^+_\ell$ (or $X^-_\ell$) has parameters $k(X) - 1 \le k(X_3) < k(X)$, $n(X_3) = n(X)$, and minimum angle $\varphi(X_3) \ge \varphi(X)$.*

**Proof.** For any spherical code $X \subset S^{n-1}$ there is a choice of $\ell$ such that

$$\frac{\operatorname{card} X}{2} \leq \operatorname{card} X_\ell^{\pm} < \operatorname{card} X. \tag{2.11}$$

Indeed, it suffices to first choose a hyperplane $L$ that dissects $X$ into two parts with different numbers of points: $\operatorname{card} X \cap \mathbf{R}^n_{\ell,+} \neq \operatorname{card} X \cap \mathbf{R}^n_{\ell,-}$. To prove the upper bound in (2.11), notice that if we had

$$\operatorname{card} X \cap \mathbf{R}^n_{\ell,\pm} < \frac{\operatorname{card} X}{2}$$

then the total number of points $\operatorname{card} X_\ell^+ + \operatorname{card} X_\ell^-$ would be smaller than $\operatorname{card} X$.

One of the spoiled codes $X_e = X_\ell^+$ or $X_\ell^-$ then has parameter $k - 1 \leq k' < k$ since

$$2^{k-1} \leq \operatorname{card} X_e < 2^k,$$

where $k = \log_2 \operatorname{card} X$. Its minimal angle satisfies $\varphi(X_3) \geq \varphi(X)$, whereas the dimension $n$ remains the same.

Notice that one could have considered a spoiling operation for spherical codes generalizing the spoiling operation $C_3$ for binary codes by intersecting the spherical code with an arbitrary hyperplane, but this operation does not allow for a good lower bound on the change of the parameter $k$, hence the generalization in terms of intersections with hemispheres is preferable.

**2.7. Numerical spoiling and controlling cones for binary codes.** In the case of binary (and $q$–ary) codes, as shown in [Man81], [ManMar11], the three spoiling operations give rise to the "numerical spoiling" producing new points in the code domain. Namely, if a code $C$ exists with parameters $[n, k, d]$ then there exist also codes with parameters:

- $[n + 1, k, d]$, by application of the appropriate first spoiling operation,
- $[n - 1, k, d - 1]$, by application of the second spoiling operation,
- $[n - 1, k', d]$, by application of the third spoiling operation (to lower $k$, possibly increasing $d$), followed by the second one (to lower $d$, decreasing $n$), and then the first one (to increase $n$ again).

We can use this numerical spoiling by studying the positions of new points with respect to the *controlling cones* in the square $(R, \delta) \in [0, 1]^2$ of code parameters.

Namely, given a point $P = (R, \delta)$, consider first the following two parametrized straight lines connecting $(R, \delta)$ respectively with $(0, 1)$ and $(1, 0)$:

$$\mathcal{L}_1(P) = \{((1+t)R, (1+t)\delta - t)\}, \qquad \mathcal{L}_2(P) = \{(1+t)R - t, (1+t)\delta\} \tag{2.12}$$

Denote by $\mathcal{I}_1(P)$ be the segment of $\mathcal{L}_1(P)$ between $(R, \delta)$ and the intersection with the $\delta = 0$ axis at $(R/(1-\delta), 0)$, and by $\mathcal{I}_2(P)$ be the segment of $\mathcal{L}_2(P)$ between $(R, \delta)$ and the intersection with the $R = 0$ axis at $(0, \delta/(1-R))$.

Now we will describe the upper, lower, left, and right controlling cones of $P$, respectively denoted by $\mathcal{C}_U(P)$, $\mathcal{C}_D(P)$, $\mathcal{C}_L(P)$, $\mathcal{C}_R(P)$.

• $\mathcal{C}_U(P)$ is bounded by $\mathcal{L}_1(P) \smallsetminus \mathcal{I}_1(P)$, $\mathcal{L}_2(P) \smallsetminus \mathcal{I}_2(P)$ and the diagonal $R + \delta = 1$.

• $\mathcal{C}_D(P)$ is bounded by $\mathcal{I}_1(P)$, $\mathcal{I}_2(P)$, and the horizontal and vertical axes between $(0, 0)$ and $(0, \delta/(1-R))$ and $(R/(1-\delta), 0)$ respectively.

• $\mathcal{C}_L(P)$ is bounded by $\mathcal{I}_2(P)$, $\mathcal{L}_1(P) \smallsetminus \mathcal{I}_1(P)$ and the vertical axis between $(0, \delta/(1-R))$ and $(0, 1)$.

• $\mathcal{C}_R(P)$ is bounded by $\mathcal{I}_1(P)$, $\mathcal{L}_2(P) \smallsetminus \mathcal{I}_2(P)$ and the vertical axis between $(R/(1-\delta))$ and $(1, 0)$.

Recall that, as it is customary in the error–correcting codes literature, vertical and horizontal axes are drawn with $\delta$ on the horizontal direction and $R$ vertically, even though one writes the code parameter coordinates as $(R, \delta)$, hence the above names of the controlling cones.

The motivation for this choice of the regions $\mathcal{C}_U(P)$, $\mathcal{C}_D(P)$, $\mathcal{C}_L(P)$, $\mathcal{C}_R(P)$ lies in the fact that, if $P = (R, \delta)$ is already a code point, then the numerical spoilings give new code points

$$P_2 = \left( \frac{n}{n-1} R - \frac{1}{n-1}, \frac{n}{n-1} \delta \right) \in \mathcal{I}_2(P)$$

(with $t = 1/(n-1)$)) by applying the third spoiling operation with $k' = k - 1$; and

$$P_1 = \left( \frac{n}{n-1} R, \frac{n}{n-1} \delta - \frac{1}{n-1} \right) \in \mathcal{I}_1(P)$$

by applying the second spoiling operation.

**2.8. Numerical spoiling for spherical codes.** We will now state an analog for spherical codes of the "numerical spoiling" of Corollary 1.2.1 of [ManMar11].

**2.8.1. Lemma.** *Assume that there exists a spherical code $X$ with* $\operatorname{card} X > 1$ *and code parameters* $[n, k, \cos\varphi]$ *where* $n \geq 2$ *is the dimension,* $X \subset S^{n-1}$ *and* $k = \log_2 \operatorname{card} X$*, with minimum angle in the small angle range* $0 \leq \varphi \leq \pi/2$*. Then there are also spherical codes in the small angle range* $0 \leq \phi \leq \pi/2$ *with parameters*

(i) $[n+1, k, \lambda\cos\varphi + 1 - \lambda]$*, for all* $\lambda \in [0, 1]$*;*

(ii) $[n-1, k, (1+u)\cos\varphi \pm u]$ *for* $u = (1 - \xi_{X,L})^2/\xi_{X,L}^2$*;*

(iii) $[n-1, k-a, \cos\varphi]$*, for any integer* $a$ *with* $0 < a < k$*.*

**Proof.** Codes with parameters (i) can be obtained by applying the first spoiling operations to $X$ with varying choices of the hyperplane $H$. In particular, $\lambda = \rho_H^2$. Codes (ii) also can be obtained directly by applying the second spoiling operation.

To obtain the third class of points, we will first explain how to obtain a spherical code with parameters $[n-1, k-1, \cos\varphi]$. Start with applying a third spoiling operation to the given code $X$ to obtain a code with parameters $[n, k-1, \cos\varphi']$ for some $\cos\varphi' \leq \cos\varphi$. Then apply the second spoiling operation twice to get a code with parameters $[n-2, k-1, \cos\varphi'']$ with $\cos\varphi'' = (1+u')((1+u)\cos\varphi' - u) - u' \leq \cos\varphi' \leq \cos\varphi$. Finally, apply the first spoiling operation for a $0 \leq \lambda \leq 1$ such that $\lambda\cos\varphi'' + 1 - \lambda = \cos\varphi$. This gives a code with parameters $[n-1, k-1, \cos\varphi]$.

More generally, to obtain a code with parameters $[n-1, k-a, \cos\varphi]$ first apply the third spoiling operation $a$ times to obtain a code with parameters $[n, k-a, \cos\varphi^a]$ with $\cos\varphi^a \leq \cos\varphi$. Then apply the second spoiling operation twice to obtain a code with parameters $[n-1, k-a, (1+u')((1+u)\cos\varphi^a - u) - u']$. Finally apply the first spoiling operation once with $\lambda$ satisfying $\lambda(1+u')((1+u)\cos\varphi^a - u) - u' + 1 - \lambda = \cos\varphi$.

**2.8.2. Remark.** According to Lemma 2.8.1 (i), if $(R, \varphi)$ is a code point, then the entire line segment

$$\ell_{n,k,\cos\varphi} = \left\{ \left( \frac{n}{n+1} R, \lambda\cos\varphi + 1 - \lambda \right) \mid \lambda \in [0, 1] \right\}$$

consists entirely of code points, and is therefore contained in $\mathcal{A}$, though it is not necessarily contained in $\mathcal{U}$, as the following example shows.

**2.8.3. Example.** In [Ran55] It is shown that, for any $\pi/2 < \varphi \leq \pi$, there exist $n$ and a spherical code $X \subset S^{n-1}$ producing the code points with

$$R(X) = \frac{1}{n}\log_2\left(\frac{\cos\varphi - 1}{\cos\varphi}\right), \text{ if } -1 \leq \cos\varphi \leq -1/n,$$

$$R(X) = \frac{1}{n}\log_2(n+1), \text{ if } -1/n \leq \cos\varphi < 0.$$

Starting with such a code point $(R(X), \cos\varphi)$ and repeatedly applying the first spoiling operation, we will obtain a sequence of segments

$$\left(\frac{n}{n+m}R(X), \lambda\cos\varphi + 1 - \lambda\right)$$

for $\lambda \in [0,1]$, which accumulate at the $R = 0$ axis for $m \to \infty$. The resulting lines are presented in the Figure 3 below, assuming that the starting point $(R, \varphi)$ lies on the curve $\frac{1}{n}\min\{\log_2(n+1), \log_2(\frac{\cos\varphi-1}{\cos\varphi})\}$ with $n = 2$.

More precisely, the figure shows the scaled curves

$$\frac{n}{n+m}\min\{\log_2(n+1), \log_2(\frac{\cos\varphi - 1}{\cos\varphi})\},$$

for $n = 2$ and $m = 1, \ldots, 5$. The segments obtained in this way belong to $\mathcal{A}$ but not $\mathcal{U}$. As these segments extend to the low angle region $0 \leq \pi \leq \pi/2$, they will encounter the $\mathcal{U}$ region for sufficiently small angles, as we will see later.

**2.8.4. Remark.** By applying the second spoiling operation to a code point $(R, \varphi)$ with varying choices of the line $\ell$, one also obtains a segment $(\frac{n}{n-1}R, (1 + u)\cos\varphi \pm u)$ for a range of possible values of $u = u(X, \ell)$ that is contained in $\mathcal{A}$, but not necessarily in $\mathcal{U}$. Similarly, when applying the third spoiling operation with varying $\ell$, one can obtain a continuous range of variability of the minimum angle $\varphi_\ell = \varphi_{X_3} \geq \varphi_X$, and a corresponding segment $\{(R - \frac{1}{n}, \phi_\ell)\}$ in $\mathcal{A}$, but not necessarily in $\mathcal{U}$.

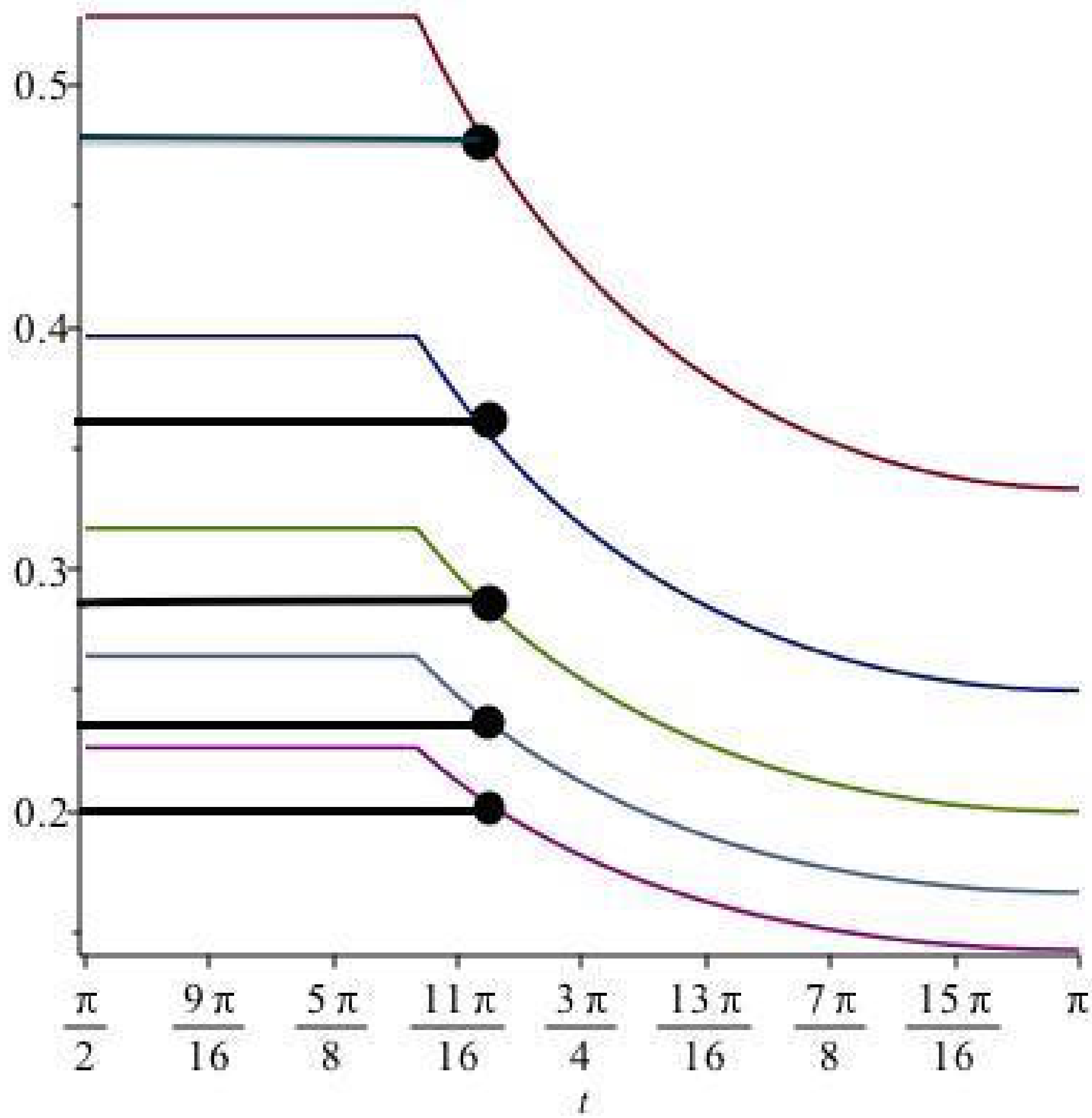

**FIG. 3**

**2.9. Existence of the asymptotic bound for spherical codes.** We show the existence of the asymptotic bound $\Gamma = \{(\alpha(\varphi), \varphi)\}$ in the small angle region $0 < \varphi \leq \pi/2$, as the boundary of the region $\mathcal{U} \subset \mathcal{A}$ of accumulation points of code points that are surrounded by an open 2-dimensional ball of accumulation points of code points, see (2.4) in the Definition 2.1.1.

**2.9.1. Lemma.** *The region $\mathcal{U}$ in the Definition 2.1.1, (2.3), for small angles $0 < \varphi \leq \pi/2$ is contained in the region $\mathcal{S}$ of (2.6), the undergraph of the function $H(\varphi)$ of (2.7).*

**Proof.** Let $P = (R, \varphi)$ be a point in $\mathcal{U}$. Since for some $\varepsilon > 0$ the ball $B(P, \varepsilon)$ is densely filled with code points, there exists a sequence $X_\ell$ of spherical codes $X_\ell \subset S^{n_\ell - 1}$ with code points $(R(X_\ell), \varphi_{X_\ell})$ converging to $(R, \varphi)$, and we can assume even that $R(X_\ell) \neq R$ for all $\ell$. For a fixed $n$ and a fixed $\varphi$ there is a bound card $X \leq M(n, \varphi)$ on the number of points of a spherical code in $S^{n-1}$ with minimum angle $\geq \varphi$. This implies that a convergent sequence $n_\ell^{-1} \log_2 \text{card } X_\ell \to R$ that is not eventually constant must have both card $X_\ell$ and $n_\ell$ unbounded. Thus, we can assume $n_\ell \to \infty$. Then the codes $X_\ell$ with large $n_\ell$ will satisfy the Kabatiansky–Levenshtein inequality, hence $R(X_\ell) \leq H(\varphi_{X_\ell})$, for $\ell \to \infty$, which implies $R \leq H(\varphi)$ for each point $(R, \varphi) \in \mathcal{U}$.

**2.9.2. Remark.** The argument of Lemma 2.9.1 does not apply to points in $\mathcal{A} \smallsetminus \mathcal{U}$, such as the points in the line segments of Example 2.8.3, for which $R(X_\ell) = \frac{n}{n+1} R(X)$ is fixed while only $\cos \varphi_{X_\ell} = \lambda_\ell \cos \varphi + 1 - \lambda_\ell$ varies. Thus, segments in $\mathcal{A} \smallsetminus \mathcal{U}$ obtained through the first spoiling operation can be found in the region above the graph $R = H(\varphi)$. Similar statements hold for segments obtained via the other spoiling operations (Remark 2.8.2), which also have fixed $R$ and varying angle.

For the purpose of this argument, it is convenient to draw the curves in the plane of coordinates $0 \leq R < \infty$ and $0 \leq \cos \varphi \leq 1$.

Due to the fact that $R$ is unbounded when the minimum angle $\varphi \to 0$, we introduce a cutoff in the region $(R, \cos \varphi)$ which we will later remove by sending the cutoff to zero. Let $\varphi_c > 0$ be a small angle and consider codes with minimum angle bounded by $\varphi \geq \varphi_c$. In this range, we consider the bound $R \leq H(\varphi_c)$, where $H(\varphi)$ is the function (2.7). Let $a_c = \lfloor H(\phi_c) \rfloor$.

**2.9.3. Lemma.** *Starting with a small $\varphi_c > 0$ and a spherical code $X$ with parameters $[n, k, \cos \varphi]$, such that $\varphi > \varphi_c$, and both $k = \log_2 \text{card } X \geq a_c$ and* *n sufficiently large, it is possible to obtain, using the spoiling operations, a new spherical code with parameters*

$$\left[n-1, k-a_c, \frac{n}{n-1}\cos\varphi - \frac{\cos\varphi_c}{n-1}\right].$$

**Proof.** The procedure is similar to the third case of the numerical spoiling of Lemma 2.8.1. We first apply the third spoiling operation, followed by two second spoiling operations, in order to obtain a code with parameters $[n-2, k-1, \cos\varphi'']$ where $\cos\varphi'' = (1+u')((1+u)\cos\varphi' - u) - u' \leq \cos\varphi'$. Assuming that the hyperplanes $L, L'$ in the second spoiling operation are such that $\cos\varphi'' < \cos\varphi$ and $n$ is so large that also $\frac{n}{n-1}\cos\varphi - \frac{\cos\varphi_c}{n-1} \geq \cos\varphi''$ , we can then apply the first spoiling operation with a parameter $\lambda \in [0,1]$ such that

$$\lambda\cos\varphi'' + 1 - \lambda = \frac{n}{n-1}\cos\varphi - \frac{\cos\varphi_c}{n-1}.$$

**2.9.4. Lemma.** *Let $X$ be a spherical code with parameters $[n, k, \cos\varphi]$. Then it is possible to obtain by spoiling another spherical code with parameters*

$$[n+1, k, \frac{n}{n+1}\cos\varphi + \frac{1}{n+1}]$$

**Proof.** This follows by using the first spoiling operation with $\lambda = n/(n+1)$.

**2.9.5. Boundaries of controlling regions.** We will describe the segments $\mathcal{R}_{L,c}(P)$, $\mathcal{R}_{R,c}(P)$, $\mathcal{R}_{U,c}(P)$, $\mathcal{R}_{D,c}(P)$, associated to a point $P = (R_0, \varphi_0)$ in the undergraph $\mathcal{S}$ in the low angle range. These controlling regions will depend on the cutoff $\varphi \geq \varphi_c$ of the region $\mathcal{Z}_c$.

Let $\mathcal{L}_1(P)$ be the line connecting the point $(R = 0, \cos\varphi = -1)$ with the point $P = (R, \theta)$. Note that $(R = 0, \cos\varphi = -1)$ is outside of the domain we are considering, since it corresponds to the large angle region $\varphi = \pi$, but we consider here only the segment of this line contained in the region

$$\mathcal{Z}_c := \{(R, \cos\varphi) \,:\, 0 \leq R \leq H(\varphi_c),\, \varphi_c \leq \varphi \leq \pi/2\}.$$

Let $\mathcal{L}_{2,c}(P)$ be the line from the point $(a_c, \cos\varphi_c)$ to the point $(R, \cos\varphi)$. We also denote by $\mathcal{I}_{1,c}(P) \subset \mathcal{L}_1(P)$ the segment of $\mathcal{L}_1(P)$ between $(R, \cos\varphi)$ and the point where it intersects the vertical axis $\cos\varphi = \cos\varphi_c$. Similarly, let $\mathcal{I}_{2,c}(P) \subset \mathcal{L}_2(P)$ be the segment of the second line between the point $(R, \cos\phi)$ and the point where it intersects the horizontal line $R = 0$. We write $\mathcal{J}_1(P) = (\mathcal{L}_1(P) \setminus \mathcal{I}_{1,c}(P)) \cap \mathcal{Z}_c$ for the complementary arc of the first line, and similarly $\mathcal{J}_{2,c} = (\mathcal{L}_2 \setminus \mathcal{I}_{2,c}) \cap \mathcal{Z}_c$.

We will now define controlling regions for spherical codes.

**2.9.6. Definition.** *a) The left controlling region $\mathcal{R}_{L,c}(P)$ is bounded by the segments $\mathcal{J}_1(P)$, $\mathcal{I}_{2,c}(P)$, and the segments of vertical axis $\cos\varphi = 0$ and horizontal axis $R = 0$ between them.*

*b) The right controlling region $\mathcal{R}_{R,c}(P)$ is bounded by the segments $\mathcal{J}_{2,c}(P)$, $\mathcal{I}_{1,c}(P)$ and the segments of the vertical line $\cos\varphi = \cos\varphi_c$ and horizontal line $R = a_c$ between them.*

*c) The lower controlling region $\mathcal{R}_{D,c}(P)$ is bounded by the segments $\mathcal{I}_{2,c}(P)$ and $\mathcal{I}_{1,c}(P)$ and the segments of the horizontal axis $R = 0$ and the vertical line $\cos\varphi = \cos\varphi_c$ between them.*

*d) The upper controlling region $\mathcal{R}_{U,c}(P)$ is bounded by the segments $\mathcal{J}_1(P)$ and $\mathcal{J}_{2,c}(P)$ and by the segments of the vertical axis $\cos\varphi = 0$ and the horizontal line $R = a_c$ between them.*

**2.9.7. Lemma.** *For any point $P = (R, \varphi)$ in $\mathcal{U}$, its lower controlling region $\mathcal{R}_{D,c}(P)$ is also contained in $\mathcal{U}$.*

**Proof.** Let $P = (R, \varphi)$ be a point in $\mathcal{U} \cap \mathcal{Z}_c$. Then there exists a sequence of code points $P_j = (R_j, \varphi_j)$ with $P_j \to P$ for $j \to \infty$. Note that this implies that $k_j \to \infty$ and $n_j \to \infty$ with their ratio converging to $R$.

Consider the region $\mathcal{R}_{D,c}(P)$. If a code point $P_j$ is sufficiently close to $P$ then the upper boundary of $\mathcal{R}_{D,c}(P)$ (the lines $\mathcal{I}_{2,c}(P)$ and $\mathcal{I}_{1,c}(P)$) is also very close to the upper boundary $\mathcal{I}_{2,c}(P_j) \cup \mathcal{I}_{1,c}(P_j)$ of $\mathcal{R}_{D,c}(P_j)$. Since each $P_j$ is a code point, there exist respective codes $X_j$. Denote their parameters by $[n_j, k_j, \cos\varphi_j]$.

By applying the spoiling operations as in Lemma 2.9.3 and Lemma 2.9.4, we obtain new codes with code points $[n_j - 1, k_j - a_c, \frac{n_j}{n_j - 1}\cos\varphi - \frac{\cos\varphi_c}{n + j - 1}]$ (assuming that $n_j$ and $k_j$ are sufficiently large) and $[n_j + 1, k_j, \frac{n_j}{n_j + 1}\cos\varphi + \frac{1}{n_j + 1}]$. These points lie, respectively, on the lines $\mathcal{I}_{2,c}(P_j)$ and $\mathcal{I}_{1,c}(P_j)$.

As we let $j \to \infty$, these points approximate points on the lines $\mathcal{I}_{2,c}(P)$ and $\mathcal{I}_{1,c}(P)$, which are therefore also in $\mathcal{U}$. By reapplying the same procedure to the obtained points on these lines, we obtain other points that densely populate nearby regions of the lines. Note moreover that, by the first spoiling operation, if points of the boundary lines $\mathcal{I}_{2,c}(P)$ and $\mathcal{I}_{1,c}(P)$ are in $\mathcal{U}$, then the entire region $\mathcal{R}_{D,c}(P)$ is also in $\mathcal{U}$.

We denote by $\Gamma_c$ the upper boundary of the region $\mathcal{U}$ inside the cutoff region $\mathcal{Z}_c$, that is,

$$\Gamma_c = \{(\alpha_c(\varphi), \varphi) \,|\, \alpha_c(\varphi) = \sup\{R \,:\, (R, \varphi) \in \mathcal{U} \cap \mathcal{Z}_c\}\,\}. \tag{2.13}$$

Given two points $P_1, P_2$ in the undergraph (2.6) $\mathcal{S}$ in $\mathcal{Z}_c$, with $P_i = (R_i, \varphi_i)$ and $\cos\varphi_1 < \cos\varphi_2$, the controlling quadrangle is the region

$$\mathcal{R}_c(P_1, P_2) = \mathcal{R}_{R,c}(P_1) \cap \mathcal{R}_{L,c}(P_2).$$

**2.9.8. Lemma.** *Given $P_1, P_2 \in \Gamma_c$, all points of $\Gamma_c$ between $P_1$ and $P_2$ belong to $\mathcal{R}_c(P_1, P_2)$.*

**Proof.** Consider two points $P_1, P_2 \in \Gamma_c$ with $\cos\varphi_1 < \cos\varphi_2$. Then $P_1 \in \mathcal{R}_{L,c}(P_2)$ and $P_2 \in \mathcal{R}_{R,c}(P_1)$. Obviously $P_1 \in \mathcal{R}_{L,c}(P_2)$ iff $P_2 \in \mathcal{R}_{R,c}(P_1)$, since if $P_2 \in \mathcal{R}_{R,c}(P_1)$ then $P_1$ is below $\mathcal{L}_1(P_2)$ and left of $\mathcal{L}_2(P_2)$ and viceversa. Similarly $P_1 \in \mathcal{R}_{U,c}(P_2)$ iff $P_2 \in \mathcal{R}_{D,c}(P_1)$. If $P_1 \notin \mathcal{R}_{L,c}(P_2)$ then it must be that $P_1 \in \mathcal{R}_{U,c}(P_2)$, but then $P_2 \in \mathcal{R}_{D,c}(P_1)$, and a point in the interior of $\mathcal{R}_{D,c}(P_1)$ would not be in $\Gamma_c$. Thus, if $P \in \Gamma_c$ is between $P_1$ and $P_2$, then $P \in \mathcal{R}_{R,c}(P_1) \cap \mathcal{R}_{L,c}(P_2)$. This proves the Lemma.

Then the same argument as was used in the case of the binary and $q$–ary codes (see [Man81], [ManMar11]) proves the following existence theorem for the asymptotic bound, as a consequence of the previous lemmata, when we let the cutoff $\varphi_c \to 0$ and $a_c \to \infty$.

**2.9.9. Theorem.** *For each $\varphi$ in the small angle range $0 \le \varphi \le \pi/2$, the set $\Gamma$ of (2.13) is the graph of a continuous monotonically decreasing function $R = \alpha(\varphi)$ with $\alpha(\varphi) \to \infty$ when $\varphi \to 0$ and $\alpha(\pi/2) = 0$. The set $\mathcal{U}$ is the undergraph of this function:*

$$\mathcal{U} = \{(R, \varphi) \,|\, R \le \alpha(\varphi)\}$$

*and is the union of all the lower controlling regions $\mathcal{R}_L(P)$ of all points $P \in \Gamma$.*

The complement of the region $\overline{\mathcal{U}} := \mathcal{U} \cup \Gamma$ then consists of two parts. One part is the remaining set of accumulation points $\mathcal{A} \smallsetminus \overline{\mathcal{U}}$, which the union of sequences of segments with fixed $R$ and varying $\cos\varphi$, resulting from spoiling operations with continuous parameters. Another part is the set $\mathcal{P} \smallsetminus \mathcal{A}$ consisting of isolated code points.

For example, code points $(R(X), \varphi_X)$ in the large angle range, producing the bound

$$\text{card } X = \frac{\cos\varphi_X - 1}{\cos\varphi_X},$$

or with card $X = n+1$, are not obtained by spoiling from other spherical codes belonging to the set $\mathcal{P} \smallsetminus \mathcal{A}$: each such point generates a sequence of segments in $\mathcal{A} \smallsetminus \overline{\mathcal{U}}$ by spoiling, as in Example 2.8.3. A characterization of the complementary set $\mathcal{P} \smallsetminus (\overline{\mathcal{U}} \cap \mathcal{P})$ is given in the next subsection in terms of multiplicities and the set of codes up to isometries that realize the code point.

**2.10. The asymptotic bound and multiplicity of code parameters.** In the case of binary or $q$–ary codes, we know that code points above the asymptotic bound are isolated and have finite multiplicity, whereas code points below the asymptotic bound form a dense set, each point of which appears with infinite multiplicity, see Theorem 2.11 of [ManMar11].

Cf. also a version of this result involving a slightly different definition of code points and avoiding appeal to the topology of $[0,1]^2$ in sec. 1.1.1.

In the case of spherical codes, the situation is different, not only because one can always apply arbitrary global isometries of the ambient sphere $S^{n-1}$ and obtain codes with the same $(n, k, \varphi)$, but because there are also spherical codes that are not rigid, namely that admit continuous deformations that are not global isometries of the ambient sphere, see [CoJiKuTo11]. Thus, we need to take these possibilities into account. First of all we only consider codes up to isometries of the ambient sphere. Following the terminology of [CoJiKuTo11], a spherical code is called "rigid" (or "jammed") if it admits no other deformations that preserve the minimum angle $\varphi$, except the global isometries.

**2.10.1. Theorem.** *A code point $P = (R, \varphi) \notin \Gamma$ lies in the region $\mathcal{U}$ if and only if it has infinite multiplicity and there exists a sequence $X_i$ of spherical codes with $(R(X_i), \varphi_{X_i}) = (R, \varphi)$ and with $n_i \to \infty$ and* card $X_i \to \infty$.

**Proof.** If $P = (R, \varphi)$ is the code point of infinitely many spherical codes $X_i$ with $n_i \to \infty$ and card $X_i \to \infty$, then by applying the spoiling operations to the codes $X_i$ with parameters $[k_i, n_i, \cos\varphi_i]$, we obtain new codes with $R_i = \frac{n_i}{n_i-1}R$ or with $R_i = \frac{n_i}{n_i+1}R$, and $\cos\varphi_i$ varying in a small interval around $\cos\varphi$, by combining the numerical spoiling operations of Lemma 2.8.1, hence a sequence of code points filling densely a surrounding ball $B(P, \varepsilon)$, for some $\varepsilon > 0$.

Conversely, if $P$ is in $\mathcal{U}$ then we will apply repeatedly the first numerical spoiling of Lemma 2.8.1 to code points in $B(P, \varepsilon)$ with $R' = R + \varepsilon'$, $\epsilon' < \epsilon$, so that $R = \frac{n}{n+m}(R + \varepsilon')$. Thus we obtain infinitely many codes $X_i$ with $k_i, n_i \to \infty$ with code point $P$.

**2.10.2. Remark.** The statement above can be rephrased in the following way: a code point $P = (R, \varphi) \notin \Gamma$ belongs to $\mathcal{P} \smallsetminus (\overline{\mathcal{U}} \cap \mathcal{P})$ if and only if either it has finite multiplicity (it is the code point of only finitely many rigid codes up to global isometries), or it has infinite multiplicity, but it is realized only by non–rigid codes with fixed $k_i = k$ and $n_i = n$ (or with at most finitely many different values of $k_i, n_i$ with $k_i/n_i = R$).

## 3. Sphere packings and the asymptotic bound.

**3.1. Spherical codes from sphere packings.** We recall briefly some of the commonly used methods for associating spherical codes to sphere packings. The bounds on code parameters of spherical codes can then be related to the density of the sphere packings, providing estimates for the maximal density. While in dealing with problems regarding the asymptotic behaviour of spherical codes one considers the length $n$ of the code as varying, and in fact one is typically interested in the behaviour for large $n \to \infty$, in the questions regarding density of sphere packings, one is typically working with a fixed dimension $n$, so these two points of view are in some sense complementary. However, as we argue below, one can investigate the location of spherical codes derived from optimal sphere packings with respect to asymptotic bounds in the space of code parameters.

**3.1.1. Spherical codes and generating functions of sphere packings.** Consider a sphere packing $\mathcal{P}$ of $\mathbf{R}^n$ by spheres $S^{n-1}_\rho$ of radius $\rho$. Choose an origin $x_0$ of the coordinates of $\mathbf{R}^n$, and choose a distance $u > 0$. Let $N = N(\mathcal{P}, u)$ be the number of centers of spheres of the packing that are at distance $u$ from $x_0$. By rescaling the coordinate vectors of the centers of these spheres, one obtains a set of $N$ points on the unit sphere $S^{n-1}$ centered at $x_0$. We denote the resulting spherical code as $X_{\mathcal{P},x_0,\rho,u}$. The minimal angle of $X_{\mathcal{P},x_0,\rho,u}$ is given by ([ConSlo99], p.26)

$$\varphi = 2\sin^{-1}\left(\frac{\rho}{u}\right).$$

The number of points $N = \text{card } X_{\mathcal{P},x_0,\rho,u}$ is determined by the theta function of the packing.

These theta functions are the generating series for the number of points in a lattice and of sphere centers in a sphere packing at a fixed distance from the origin. More precisely, they are defined as follows ([ConSlo79], Chapter 2, Sec. 2.3).

Let $\Lambda \subset \mathbf{R}^n$ be a lattice, and let $N_\Lambda(m)$ denote the number of points $x \in \Lambda$ such that $\langle x, x\rangle = m$. Then one sets

$$\Theta_\Lambda(z) = \sum_{x\in\Lambda} q^{\langle x,x\rangle} = \sum_{m=0}^{\infty} N_\Lambda(m)\, q^m,$$

with $q = e^{\pi i z}$.

Let now $\mathcal{P}$ be a periodic sphere packing of $\mathbf{R}^n$, where the centers of the spheres are placed at a finite number of translates of a lattice $\Lambda$, that is at the set of points of the form

$$u_j + \Lambda, \quad j = 1, \dots, \ell,$$

with the vectors $u_j$ such that $u_j - u_k \notin \Lambda$, for $j \neq k$. Then one sets

$$\Theta_{\mathcal{P}}(z) = \frac{1}{\ell}\sum_{j=1}^{\ell}\sum_{k=1}^{\ell}\sum_{x\in\Lambda} q^{\langle x+u_j-u_k, x+u_j-u_k\rangle}.$$

**3.1.2. Sphere packings and kissing configurations.** A spherical code can be thought of as a packing of non-overlapping spherical caps on the surface of an $S^{n-1}$ sphere with the code points as the centers of the caps. One can obtain a spherical code $X \subset S^{n-1}$ with minimum angle $\varphi \geq \pi/3$ from a sphere packing $\mathcal{P}$ of $\mathbf{R}^n$ by considering the points of tangency between adjacent spheres in the packing, *the kissing configuration.* Lower bounds on the kissing numbers can be obtained by direct construction of such spherical codes while upper bounds are obtained via estimates on the maximum number $M(n, \varphi)$ of code points on $S^{n-1}$ with minimum angle $\varphi$.

It is shown in [KaLe78] that the bound $M(n, \varphi)$ on the maximum number of points of a spherical code $X \subset S^{n-1}$ with minimum angle $\varphi$ also provides an upper bound for the minimal density $\Delta_n$ of a sphere packing in $\mathbf{R}^n$, by

$$\Delta_n \leq \sin^n(\varphi/2) M(n+1, \varphi). \tag{3.1}$$

This bound holds for all $0 < \varphi \leq \pi$ since passing to a higher dimension $n+1$ makes it possible to lower the angle (as in the spoiling operations discussed above). An improved bound obtained in [CoZh14] shows that for $\pi/3 \leq \theta \leq \pi$ (the minimum angle range of sphere packings) one has

$$\Delta_n \leq \sin^n(\varphi/2) M(n, \varphi). \tag{3.2}$$

This bound is obtained by considering a sphere $S^{n-1}_R$ of radius $R \geq 2$ that contains at least $\Delta R^n$ sphere centers, where $\Delta$ is the density of the packing, and such that the center of $S^{n-1}_R$ is not one of them, and projecting these points from the center onto the surface of $S^{n-1}_R$. The resulting spherical code has minimum satisfying $\sin(\theta/2) = 1/R$, and number of points $\Delta R^n \leq M(n, \theta)$, showing (3.2), see Proposition 2.1 of [CoZh14].

**3.1.3. Wrapped spherical codes from sphere packings.** We describe one more construction associating spherical codes to sphere packings, which provides a family of spherical codes whose asymptotic density approaches the density of the sphere packing, see [HamZeg97]. This method will be useful to relate sphere packings of maximal density to asymptotic bounds in the space of spherical codes.

Unlike the previous constructions that relate sphere packings in $\mathbf{R}^n$ to spherical codes in $S^{n-1}$, in [HamZeg97] one constructs "wrapped" spherical codes in $S^{n-1}$ from sphere packings in $\mathbf{R}^{n-1}$. The construction is based on partitioning the sphere $S^{n-1}$ into annuli

$$A_i = \{x = (x_1, \dots, x_n) \in S^{n-1} \,|\, \alpha_i \leq \sin^{-1}(x_n) \leq \alpha_{i+1}\},$$

with latitudes $-\pi/2 = \alpha_0 < \cdots < \alpha_N = \pi/2$ for some $N$ sufficiently large, and using low distortion maps between the annuli $A_i$ and regions $U_i \subset \mathbf{R}^{n-1}$, in order to map increasingly large regions of $\mathbf{R}^{n-1}$ to the sphere $S^{n-1}$ in such a way that the density of the packing becomes sufficiently close to the code density.

More precisely, the code density $\Delta_X$ of a spherical code $X \subset S^{n-1}$ is the fraction of the sphere $(n-1)$–dimensional area that is covered by the disjoint spherical caps associated to the spherical code:

$$\Delta_X = \text{card } X \cdot \frac{S(n,\varphi)}{S_n}.$$

Here $S_n = \frac{n\pi^{n/2}}{\Gamma(\frac{n}{2}+1)}$ is the $(n-1)$-dimensional area of the sphere $S^{n-1}$ and

$$S(n,\varphi) = S_{n-1} \int_0^{\varphi/2} \sin^{n-2}(x)\, dx.$$

The maximum possible density of a spherical code $X \subset S^{n-1}$ for a fixed $n$ is then given by

$$\Delta(n,\varphi) = M(n,\varphi)\frac{S(n,\varphi)}{S_n},$$

hence it can be estimated on the basis of estimates on the maximum number $M(n,\varphi)$ of points of a spherical code $X \subset S^{n-1}$ with minimum angle $\varphi$. One also defines

$$\Delta_n^c = \lim_{\varphi \to 0} \Delta(n,\varphi). \tag{3.3}$$

A family of spherical codes $X_\ell \subset S^{n-1}$ is asymptotically optimal if

$$\lim_{\ell\to\infty} \frac{\text{card } X_\ell}{M(n,\varphi_\ell)} = 1$$

where the minimum angle $\varphi_\ell$ of the code $X_\ell$ satisfies $\varphi_\ell \to 0$ for $\ell \to \infty$, hence

$$\lim_{\ell\to\infty} \frac{\Delta_{X_\ell}}{\Delta(n,\varphi_\ell)} = 1.$$

If $\Delta_n^P$ denotes the maximal density of sphere packings in $\mathbf{R}^n$, then it is known that

$$\Delta_n^c = \Delta_{n-1}^P. \tag{3.4}$$

Thus, it is possible to approximate the maximal density of sphere packings using a family of asymptotically optimal spherical codes. It is shown in [HamZeg97] that the wrapped spherical codes obtained from an optimal sphere packing are such an asymptotically optimal family, if the latitudes of the annuli are chosen with the condition that $\max_i \delta\alpha_i + 2\sin(\varphi/2)\min_i \delta\alpha_i \to 0$ for $\varphi \to 0$, with $\delta\alpha_i = \alpha_{i+1} - \alpha_i$.

**3.2. Sphere packings and the space of spherical code parameters.** In the space of code parameters of spherical codes, one considers spherical codes in arbitrary dimension $n$, while in the study of densities of spherical packings one is interested in the maximal density for a fixed $n$. The maximal density $\Delta_n^P$ decays exponentially with $n \to \infty$, as one can see from the estimates in [KaLe78].

We first show that isolated code points above the asymptotic bound for spherical codes can be used to construct asymptotically optimal sequences of spherical codes in $S^{n-1}$ whose densities approximate the maximal density for sphere packings in $\mathbf{R}^{n-1}$. We then show that the wrapped spherical codes in $S^{n-1}$ obtained from maximal density sphere packings provide a construction of code points that are close to the asymptotic bound of spherical codes or above it.

Start with a sequence $\varphi_{c,\ell}$ of cutoff angles $\varphi_{c,\ell} \to 0$ as $\ell \to \infty$ and consider the intervals $\mathcal{I}_\ell = [\varphi_{c,\ell+1}, \varphi_{c,\ell}]$. Within each interval, consider the isolated code points that lie above the asymptotic bound $R = \alpha(\varphi)$. We consider only the isolated code points in $\mathcal{P} \smallsetminus \mathcal{A}$ rather than all the code points in $\mathcal{A} \smallsetminus \mathcal{U}$ since the lines in $\mathcal{A} \smallsetminus \mathcal{U}$ are obtained, as we saw, by spoiling of codes by embedding into higher dimensional spheres, and each such sequence of lines is dominated by an isolated point with better code parameters. Let $\mathcal{P}_\ell$ be the subset of isolated code points in $\mathcal{P} \smallsetminus \mathcal{A}$ with minimum angle $\varphi \in \mathcal{I}_\ell$.

**3.2.1. Proposition.** *For a given natural $n$, denote by $\mathcal{P}_{\ell,n} \subset \mathcal{P}_\ell$ be the set of code points realized by some code $X \subset S^{n-1}$. There is an asymptotically optimal sequence of codes $X_{\ell,j}$ realising the code points in $\mathcal{P}_{\ell,n}$ with maximal $R$ coordinate.*

**Proof.** By Theorem 2.10.1 and Remark 2.10.2, we know that all the code points $P = (R, \varphi)$ in $\mathcal{P}_\ell$ either have finite multiplicity and are points of a finite set of rigid spherical codes, or else of an infinite set of non–rigid spherical codes with bounded $k$ and $n$. Given a fixed value of $n$, consider the subset $\mathcal{P}_{\ell,n}$ of such isolated code points $P$ in $\mathcal{P}_\ell$ that there exists at least one spherical code $X \subset S^{n-1}$ producing the code point, $P = (R(X), \varphi_X)$.

The set $\mathcal{P}_{\ell,n}$ is finite because it consists of isolated points and is contained in the bounded region $\varphi \in \mathcal{I}_\ell$ and $R \le n^{-1} \log_2 M(n, \varphi)$.

If a point $P \in \mathcal{P}_{\ell,n}$ corresponds to an infinite number of non–rigid spherical codes, then the code density can be optimized over the possible representatives up to global isometries, by locally modifying the code in order to increase density, as discussed in [CoJiKuTo11].

If a point $P \in \mathcal{P}_{\ell,n}$ has finite multiplicity then the number of codes realizing these code parameters with fixed length $n$ is finite and the number of code points is also fixed by $R = n^{-1} \log_2 \operatorname{card} X$.

Consider the point (or family of such points) $P$ in the finite set $\mathcal{P}_{\ell,n}$ with the largest $R = k/n$ coordinate. To each such point we associate a finite set of spherical codes $X_{\ell,j}$ with $j = 1, \dots , N_\ell$.

If the code point has finite multiplicity, the set is given by the union of all codes with length $n$ and largest $k$ in $\mathcal{P}_{\ell,n}$ realizing the code point. If the code point has infinite multiplicity, it is given by a set of maximal density representatives of the code point among non–rigid spherical codes of length $n$ with largest $k$ in $\mathcal{P}_{\ell,n}$. This provides a sequence $X_{\ell,j}$ of spherical codes that satisfies the asymptotically optimal property, hence their code densities approximate the maximal sphere packing density.

Note that, unlike the case of wrapped spherical codes, the asymptotically optimal sequence of spherical codes constructed in this way does not come from a single sphere packing.

When we consider a sphere packing $\mathcal{P}$ in $\mathbf{R}^{n-1}$ with maximal density, we can obtain from it, through the wrapped spherical codes construction, spherical codes $X \subset S^{n-1}$ that lie on or above the asymptotic bound.

**3.2.2. Lemma.** *Let $\mathcal{P}$ be a sphere packing in $\mathbf{R}^{n-1}$ that realizes the maximal density $\Delta_{\mathcal{P}} = \Delta^P_{n-1}$. Let $X_\ell \subset S^{n-1}$ be an asymptotically optimal family of wrapped spherical codes associated to $\mathcal{P}$. For $\ell$ sufficiently large, the code points $P_\ell = (R(X_\ell), \varphi_{X_\ell})$ either lie above the asymptotic bound or approach the asymptotic bound from below.*

**Proof.** As in [HamZeg97], we construct wrapped spherical codes $X_\ell$ from the sphere packing $\mathcal{P}$ with a choice of latitude angles for the annuli such that the family $\{X_\ell\}$ is asymptotically optimal. This means that $\Delta_{X_\ell}/\Delta_{\mathcal{P}} \to 1$, or equivalently, $\operatorname{card} X_\ell \cdot M(n, \varphi_{X_\ell})^{-1} \to 1$.

Suppose that there exists an $\varepsilon > 0$ such that, for all sufficiently big $\ell \geq \ell_0$ the points $P_\ell$ that are contained in $\mathcal{U}$, remain at a distance at least $\varepsilon$ from the asymptotic bound, that is,

$$\alpha(\varphi_{X_\ell}) - R(X_\ell) \geq \varepsilon,$$

where $R = \alpha(\varphi)$ is the asymptotic bound. Then there exists a ball $B(P_\ell, \varepsilon) \subset \overline{\mathcal{U}}$. In particular, for some $0 < \varepsilon' < \varepsilon$, there exists a spherical code $X'_\ell$ with $R(X'_\ell) = R(X_\ell) + \varepsilon'$ and $\varphi_{X'_\ell} = \varphi_{X_\ell}$. This follows by applying the numerical spoiling of Lemma 2.8.1 and arguing as in Theorem 2.10.1. This then implies that

$$n^{-1} \log_2 A(n, \varphi_{X_\ell}) - R(X_\ell) \geq \varepsilon'$$

hence the $X_\ell$ would not be asymptotically optimal.